\documentclass[aip,jcp,amsmath,amssymb,longbibliography,reprint]{revtex4-1}

\usepackage{graphicx}
\usepackage{dcolumn}
\usepackage{bm}

\usepackage[utf8]{inputenc}
\usepackage[T1]{fontenc}
\usepackage{mathptmx}
\usepackage{etoolbox}
\usepackage{siunitx}
\usepackage{chemformula}
\usepackage{braket}
\DeclareSIUnit[number-unit-product = {\,}]
\cal{cal}

\makeatletter
\def\@email#1#2{%
 \endgroup
 \patchcmd{\titleblock@produce}
  {\frontmatter@RRAPformat}
  {\frontmatter@RRAPformat{\produce@RRAP{*#1\href{mailto:#2}{#2}}}\frontmatter@RRAPformat}
  {}{}
}%
\makeatother
\begin{document}


\title{Optimized basis sets for DMRG calculations of quantum chains of rotating water molecules}
\author{Tobias Serwatka}
\author{Pierre-Nicholas Roy}%
 \email{pnroy@uwaterloo.ca.}
\affiliation{ 
Department of Chemistry, University of Waterloo, Waterloo, Ontario N2L 3G1, Canada
}


\date{\today}

\begin{abstract}
In this contribution, we employ a density matrix based optimization procedure to obtain customized basis functions to describe chains of rotating water molecules in interaction regimes associated with different intermolecular distances. 
This procedure is shown to yield a very compact basis with a clear truncation criterion based on the population of the single particle basis functions. For the water trimer, we discuss the convergence behavior of several properties and show it to be superior when compared to an energy-based truncated basis. It is demonstrated that the optimized basis reduces the necessary number of basis functions by at least an order of magnitude. Finally, the optimization procedure is employed to study larger chains of up to ten water molecules. The formation of hydrogen bonds as well as its impact on the net polarization of the chain is discussed.
\end{abstract}

\maketitle


\section{Introduction}


Nanoconfined water systems are of special interest due to the unusual behaviour they can exhibit compared to bulk water~\cite{wang2021abnormal}. For instance, experimental studies of water enclosed in beryl and cordierite crystals suggest quantum-critical behaviour 
at low temperatures due to the purely dipolar interaction of the molecules.\cite{belyanchikov2020dielectric,belyanchikov2022fingerprints,gorshunov2022effect,chan2022effect} 
For water chains at large water-water distances, we
have predicted
the occurrence of a quantum phase transition between a disordered and a dipole-ordered phase.\cite{serwatka2023quantum} In a subsequent theoretical study, we proposed the appearance of ferroelectric phases in water chains in carbon nanotubes.\cite{serwatka2022ferroelectric}
The quantum mechanical description of these extended systems is challenging due to the exponential scaling of the Hilbert space with system size. 
The latter limits the use of exact diagonalization to systems composed of a dozen weakly interacting molecules.\cite{felker2017accurate,halverson2018quantifying} 
In previous work, we developed and employed a density matrix renormalization group (DMRG)\cite{white1992density} approach to study chains of dozens and hundreds of rotating molecules.
\cite{iouchtchenko2018ground,mainali2021comparison,serwatka2022ground,serwatka2022ferroelectric,serwatka2023quantum} 
In the DMRG approach, the wave function of a chain of $N$ molecules scales like $\mathcal{O}(N\cdot D\cdot d_{\mathrm{max}}^2)$ with the local Hilbert space dimension $D$ and the maximal bond dimension $d_{\mathrm{max}}$. 
The maximal bond dimension $d_{\mathrm{max}}$ is determined by the entanglement between different parts of the system.
Therefore, $d_{\mathrm{max}}$ is  mostly a property of the physical system. In contrast, the dimension of the local Hilbert space $D$, necessary to converge the wave function, strongly depends on the choice of local basis. 
In one- and two-site DMRG optimization approaches, a scaling of $\mathcal{O}(D^2)$ and $\mathcal{O}(D^3)$ is observed for given $N$ and maximal bond dimension.\cite{schollwock2011density} For some systems, like the prototypical Ising chains, the local Hilbert space is finite so $D$ is fixed (and small). But bosonic systems such as the Bose-Hubbard model or general Hamiltonians have an infinite local Hilbert space. This means that in a practical calculation, a truncated basis has to be used. 
For these systems, it is crucial to choose a compact basis that reduces the necessary basis set size and thus the numerical effort. In chemical physics, when computing ro-vibrational energy levels, one usually adopts an energy truncation criterion to optimize a basis set. This is the case in the sequential diagonalization truncation approach\cite{bavcic1988localized,bavcic1989theoretical} as applied to dipolar assemblies.\cite{felker2017accurate} 
Basis contraction techniques,\cite{bramley1994calculation,tremblay2006calculating}, or ``polyad'' truncation schemes for vibrational\cite{HALVERSON201537} and rotational\cite{halverson2018quantifying} degrees of freedom are also based on an energy truncation criterion. 

In previous work on linear water chains,\cite{serwatka2022ground,serwatka2022ferroelectric,serwatka2023quantum} we used the energy eigenstates of a single molecule as a local basis. For water, this corresponds to the eigenbasis of the asymmetric top
\begin{align}
\Ket{J_{K_{a}K_{c}}m} = \sum_{k}A_{k}^{Jm}\Ket{Jmk},\quad m=-J...,J
\label{eq:asymtop}
\end{align}
which is a linear combination of the Wigner functions $\Ket{Jmk}$. 
Note that the sum in Eq.~\eqref{eq:asymtop} is restricted to even or odd $k$ when \emph{para}- or \emph{ortho}-water is considered. 
However, the use of the asymmetric top basis has two weaknesses. First, this basis performs best in parameter regimes with small angular localizations, i.e., at large water-water distances $R$ where the intermolecular interactions are weak. 
At these distances, the system behaves perturbatively and only a small number of basis functions is needed. 
This basis choice also works in the purely dipolar region up to $R\sim\SI{6}{\angstrom}$ because single molecules are mainly aligned along the chain axis and the low-lying Wigner functions are sufficient to describe this localization.\cite{serwatka2023quantum} 
But at shorter distances, hydrogen bonds are formed and this leads to tilted water molecules with respect to the chain axis and a strong localization along all Euler angles. In this region, hundreds of rather delocalized Wigner basis functions are necessary to converge properties. For systems with more than a few molecules, such a large basis is out of reach. 
Besides, the hydrogen bonding region is also characterized by an increased entanglement. As a consequence, the maximum bond dimension, $d_{\mathrm{max}}$, has to be chosen rather large which further raises the computational cost.
A second weakness of the asymmetric top basis is the increasing size of the number of levels within the basis. Since the basis functions are derived from a system with a higher symmetry than that of the target system, the level structure within the basis has a large and increasing degeneracy pattern. 
Hence, instead of increasing the number of basis functions by a small arbitrary number, one has to increase it stepwise with increasing level size. 
For instance, when going from a \emph{para} basis with $j_{\mathrm{max}}=1$ to the next possible level size $j_{\mathrm{max}}=2$, one has to include 15 additional functions, and going to the next step from $j_{\mathrm{max}}=2$ to $j_{\mathrm{max}}=3$ already requires 21 additional functions. 
Due to these weaknesses, the asymmetric top as well as the Wigner representations do not lead to a compact basis to efficiently obtain converged or nearly converged results over the whole water-water distance range. For states represented as matrix products, different strategies, such as using pseudosites~\cite{jeckelmann1998density} or projected purified two-site DMRG~\cite{kohler2021efficient} can be used to reduce the local Hilbert spaces~.\cite{stolpp2021comparative} Instead of describing the whole $R$-space using a uniform basis derived from the non-interacting case, we employ a basis optimized at every $R$. Such a basis was proposed by Zhang, Jeckelmann and White for a chain of translationally invariant systems~.\cite{zhang1998density} The idea behind this basis is similar to the basic idea of DMRG: instead of using an energy-based basis truncation, the basis set is derived from reduced density matrices which leads to a basis that is adapted to the specific system. In the present work, we follow a different strategy to describe the whole interaction region of the water chain. The remainder of this paper is structured as follows: in Sec.~\ref{sec:theory} we briefly explain the DMRG method and the optimization procedure to generate the basis sets. Computational details are presented in Sec.~\ref{sec:comp} followed by a discussion of the results in Sec.~\ref{sec:results}. Concluding remarks are presented in Sec.~\ref{sec:conclusio}. 

\section{Theory}\label{sec:theory}
A linear chain of $N$ rotating asymmetric top molecules with equidistant fixed centers-of mass at distance $R$ has the following Hamiltonian:
\begin{align}
\hat{H} = \sum_{i=1}^{N}\left[A_{e}\hat{J}^{2}_{a,i}+B_{e}\hat{J}^{2}_{b,i}+C_{e}\hat{J}^{2}_{c,i}\right]+\sum_{<ij>}^{N}\hat{V}(\Omega_{i},\Omega_{j};R)~,
\label{eq:Hamiltonian}
\end{align}
where the kinetic energy contains the components of the angular momentum operator $\hat{\vec{J}}$ in the molecule-fixed frame as well as the rotational constants $A_{e}$, $B_{e}$ and $C_{e}$. In Eq.~\eqref{eq:Hamiltonian}, we approximate the interaction between the molecules by a sum over nearest-neighbor pair-potentials. These interactions depend on the orientations of the rotors characterized by the Euler angles $\Omega_{i}=\lbrace\theta_{i},\varphi_{i},\chi_{i}\rbrace$. The chain axis is defined as parallel to the space-fixed $z$-axis and the $b$-axis as the $z$-axis in the molecule-fixed frame.

\subsection{Density-matrix renormalization group method}
In general, the quantum states of these chains can be expanded in a chosen tensor product basis. Thus, for a chain of $N$ sites with $D$-dimensional local basis sets $\lbrace\Ket{\sigma_{i}}\rbrace$ a general quantum state can be expressed as
\begin{align}
\Ket{\Psi}=\sum_{\sigma_{1},...,\sigma_{N}}C_{\sigma_{1},...,\sigma_{N}}
\bigotimes_{i=1}^N \Ket{\sigma_{i}}.
\label{eq:expansion}
\end{align}
The exponential scaling with respect to system size $N$, of that ansatz strongly limits the applicability to longer chains. Alternatively, the quantum state can be expressed as a matrix product state (MPS) by rewriting the expansion tensor in Eq.~\eqref{eq:expansion} as 
\begin{align}
C_{\sigma_{1}...\sigma_{N}}=&\sum_{\alpha_{1},...,\alpha_{N}}M_{1\alpha_{1}}^{\sigma_{1}}\cdots M_{\alpha_{N}1}^{\sigma_{N}} \nonumber\\
=&\mathbf{M}^{\sigma_{1}}\cdots\mathbf{M}^{\sigma_{N}}.
\label{eq:mps}
\end{align}
The above form holds for open-boundary conditions. The $\mathbf{M}^{\sigma}$ are rank-3 tensors which can be regarded as matrices with an additional index $\sigma_{i}$ that labels the basis set associated with a site $i$. 
The other two matrix dimensions are called bond dimensions. 
As mentioned earlier, the size of the MPS scales as $\mathcal{O}(N\cdot D\cdot d_{\mathrm{max}}^2)$, where $d_{\mathrm{max}}$ is the maximum bond dimension of the MPS. 
If all bond dimensions are chosen such that the equality in Eq.\eqref{eq:mps} is strictly obeyed, the MPS ansatz scales as unfavorably as the general expansion in Eq.~\eqref{eq:expansion}. 
However, for one-dimensional gapped systems with short-range interactions, it is possible to use much smaller bond dimensions and still obtain a very accurate representation of the quantum state.\cite{schollwock2011density} 
The DMRG description uses such an MPS ansatz. 
Calculating the ground state in DMRG is linked to forming the Schmidt decomposition 
\begin{align}
\Ket{\Psi} = \sum_{i}\sqrt{\lambda_{i}}\Ket{\phi_{i}^{A}}\otimes\Ket{\phi_{i}^{B}},
\label{eq:schmidt}
\end{align}
where the whole system is partitioned into an $A$ part and a  $B$ part with orthonormal sets $\lbrace\Ket{\phi_{i}^{A}}\rbrace$ and $\lbrace\Ket{\phi_{i}^{B}}\rbrace$. 
During the DMRG optimization, the boundary of that bipartition is constantly shifted through the chain (sweeping) and the matrices in Eq.~\eqref{eq:mps} are optimized until convergence is reached. 
In order to keep the bond dimensions tractable, only a subset of the states in Eq.~\eqref{eq:schmidt} is kept while a threshold $\varepsilon\geq 1-\sum_{k}\lambda_{k}$ is used to define these states. 
Restricting the bond dimension defines an upper bound to the entanglement between part $A$ and part $B$. 
A standard measure of such bipartite entanglement is the von-Neumann entanglement entropy
\begin{align}
S_{\mathrm{vN}} = -\mathrm{tr}\left(\hat{\rho}_{A}\ln\hat{\rho}_{A}\right),
\end{align}
where $\hat{\rho}_{A}$ denotes the reduced density operator of subsystem $A$. Since the coefficients in the Schmidt decomposition are the eigenvalues of this reduced density operator, the von-Neumann entanglement entropy can also be formulated as
\begin{align}
S_{\mathrm{vN}} = -\sum_{i}\lambda_{i}\ln\left(\lambda_{i}\right).
\label{eq:SvN}
\end{align} 
This reveals that the entanglement entropy is intimately linked to the structure of the wave function.

\subsection{Optimized basis set}\label{sec:Obasis}
\begin{figure*}[htbp]
\centering
\includegraphics[width=\textwidth]{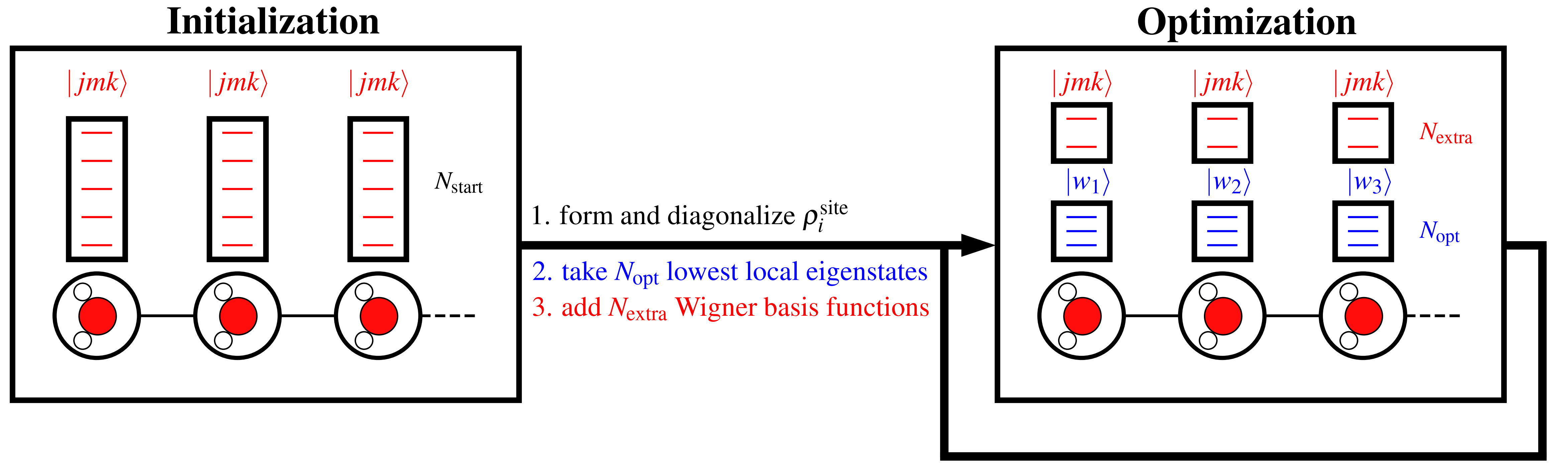}
\caption{Schematic representation of the basis optimization procedure as applied to the water chain.}
\label{fig:scheme}
\end{figure*}

We employ an iterative approach proposed by Zhang, Jeckelmann and White to construct  optimized basis functions for the rotor chain.\cite{zhang1998density} 
The procedure used to generate the optimal basis set is depicted in Fig.~\ref{fig:scheme}.  
In the first step (initialization), the ground state of the whole system is calculated in a small local basis to create a so-called ``start'' wave function 
\begin{align}
\Ket{\Psi_{\mathrm{start}}}=\sum_{\sigma_1,...,\sigma_{N}}^{N_{\mathrm{start}}}C^{\mathrm{start}}_{\sigma_{1},...,\sigma_{N}}\bigotimes_{i=1}^N \Ket{\sigma_{i}},
\label{eq:startWF}
\end{align}
where the coefficient tensor $C^{\mathrm{start}}_{\sigma_{1},...,\sigma_{N}}$ is expressed as an MPS of the form given in Eq.~\eqref{eq:mps}. In the present work, the primitive basis set $\lbrace\Ket{\sigma}\rbrace$ labels the first $N_{\mathrm{start}}$ Wigner functions. 
The wave function in Eq.~\eqref{eq:startWF} is used to generate the initial version of the new basis. 
To do so, the reduced density matrix at every site $i$ is first formed,
\begin{align}
\hat{\rho}_{\mathrm{red}}^{i} = \mathrm{tr}_{j\neq i}\Ket{\Psi_{\mathrm{start}}}\Bra{\Psi_{\mathrm{start}}},
\label{eq:redmatrix}
\end{align}   
where the trace over all remaining sites $j\neq i$ is formed. 
In a second step, all of these $N$ reduced density matrices are diagonalized. 
This leads to a set of eigenvalues $\lbrace\omega^{i}_{w}\rbrace_{w=1,..,N_{\mathrm{start}}}$ and eigenstates $\lbrace\Ket{w_{i}}\rbrace_{w=1,..,N_{\mathrm{start}}}$ for each site $i$. 
These basis sets are the first version of the new basis. 
Since they are derived from a density matrix, every single of these $N$ basis sets obeys $\sum_{w}^{N_{\mathrm{start}}}\omega_{w}^{i}=1$. In the following, the states are ordered as $\omega^{i}_{1}\geq\omega^{i}_{2}\geq ...\geq\omega^{i}_{N_{\mathrm{start}}}$. 
The larger the eigenvalues $\omega_{w}^{i}$, the more important the corresponding basis functions in a basis set expansion such as the one in  Eq.~\eqref{eq:expansion}. 
Thus, for a given primitive basis $\lbrace\Ket{\sigma}\rbrace$, 
the first $K$ functions of each local basis set $\lbrace\Ket{w_{i}}\rbrace_{w=1,..,N_{\mathrm{start}}}$ are the $K$ optimal basis functions. 
The latter is  fundamentally different from an energy-based basis truncation for which the $K$ dominating basis functions can be distributed over the whole basis set. 
Since the one-site reduced density matrices $\rho_{\mathrm{red}}^{i}$ are derived from the density of the whole system, the interactions between sites are already incorporated into the basis. 
At this stage however, the basis is only optimal for a given $j_{\mathrm{max}}$ that differs of course from the optimal basis one would obtain from the true, converged ground state. 
To obtain that basis, it is necessary to mix in more primitive functions, the higher Wigner states, into the basis. 
For that, we follow the iterative approach of Zhang \emph{et al.} which yields the optimal basis for higher $j_{\mathrm{max}}$ without generating large Hilbert spaces.\cite{zhang1998density} 
That optimization scheme is depicted in the second box of Fig.~\ref{fig:scheme}. 
At first, one has to define $N_{\mathrm{opt}}$, the number of density-based basis functions for every site which shall be optimized. 
Next, one defines $N_{\mathrm{extra}}$, the number of additional Wigner functions that are added at every optimization step. 
The scanning through the Wigner functions does not start at $j_{\mathrm{max}}=0$ but at the first Wigner functions not included in the initial wave function $\Ket{\Psi_{\mathrm{start}}}$. 
$N_{\mathrm{extra}}$ is constant at every step but the corresponding functions change since one goes through all Wigner functions for a given $j_{\mathrm{max}}$ during one optimization cycle.
At every optimization step, the $N_{\mathrm{opt}}$ density-based functions and the $N_{\mathrm{extra}}$ additional Wigner states are orthogonalized and the ground state of the chain is calculated in this orthogonalized basis. 
From that new ground state, a new set of density-basis functions is generated. The $N_{\mathrm{opt}}$ lowest of these functions are taken together with the next $N_{\mathrm{extra}}$ Wigner functions and the optimization procedure is repeated until every Wigner function is included once in the extra basis. The latter defines one iteration. 
By performing as many iterations as needed to converge the total energy of the system, this procedure leads to the $N_{\mathrm{opt}}$ optimal density-based functions for a given $j_{\mathrm{max}}$. Note that the procedure is independent of DMRG because it just aims at finding the most compact expansion tensor in Eq.~\eqref{eq:expansion}. 
In principle, this whole density-based basis optimization procedure could also be combined with an exact diagonalization technique.  For large systems however, the favourable scaling of DMRG makes that approach preferable.      

\section{Computational details}\label{sec:comp}
All calculations are performed with the q-TIP4P model,\cite{abascal2005general,habershon2009competing} 
where the OH-bond length and the bond angle are fixed at \SI{0.9419}{\angstrom} and \SI{107.4}{\degree}. 
For the rotational constants, values of $A_{e}=\SI{27.88}{\per\centi\metre}$, $B_{e}=\SI{14.51}{\per\centi\metre}$ and $C_{e}=\SI{9.29}{\per\centi\metre}$ are employed.\cite{hall1967pure} 
The density-based basis functions are expanded in a primitive Wigner basis with $j_{\mathrm{max}}=9$. In all calculations, an electric field $\vec{F}$ with $|\vec{F}\vec{\mu}|=\SI{2}{\per\centi\metre}$ is applied at the edge water molecules to break the inversion symmetry and to ensure a non-degenerate ground state. $\vec{\mu}$ denotes the electric dipole moment of the water molecule. For the trimer, we initiate the optimization procedure with $N_{\mathrm{start}}=40$ and use an extra space of $N_{\mathrm{extra}}=10$. The calculations are converged after 15 iterations at maximum. A larger initial basis of $N_{\mathrm{start}}=85$, a optimized basis size $N_{\mathrm{opt}}=50$ and an extra space of $N_{\mathrm{extra}}=10$ is employed for the chains with $N>3$. Since we only consider \emph{para} water the primitive basis is restricted to Wigner functions with even $k$. DMRG calculations are performed using the ITensor package.\cite{fishman2022itensor}

\section{Results and Discussion}\label{sec:results}
To illustrate the present basis optimization approach, we first performed ground state calculations for the water trimer. We then extended the approach to longer water chains. In all calculations, a small electric field is applied at the two edge molecules in order to break the degeneracy of the ground state. 
This is done for two reasons. First, this symmetry breaking simulates a more realistic situation because any anisotropic environment in an experimental realization of these chains would lift that degeneracy. 
Second, the DMRG optimization is variational and tends towards a state with the smallest energy. 
There is no additional constraint to obtain a specific linear combination of the two-fold degenerate ground state. Hence, it is not possible to benchmark properties like entanglement entropies or spatial distributions because for different calculations the optimizer will converge to different linear combinations. By breaking the ground state degeneracy, this issue can be avoided and unique values for those properties can be obtained.

\subsection{Water trimer}
\begin{figure}
\centering
\includegraphics[width=\columnwidth]{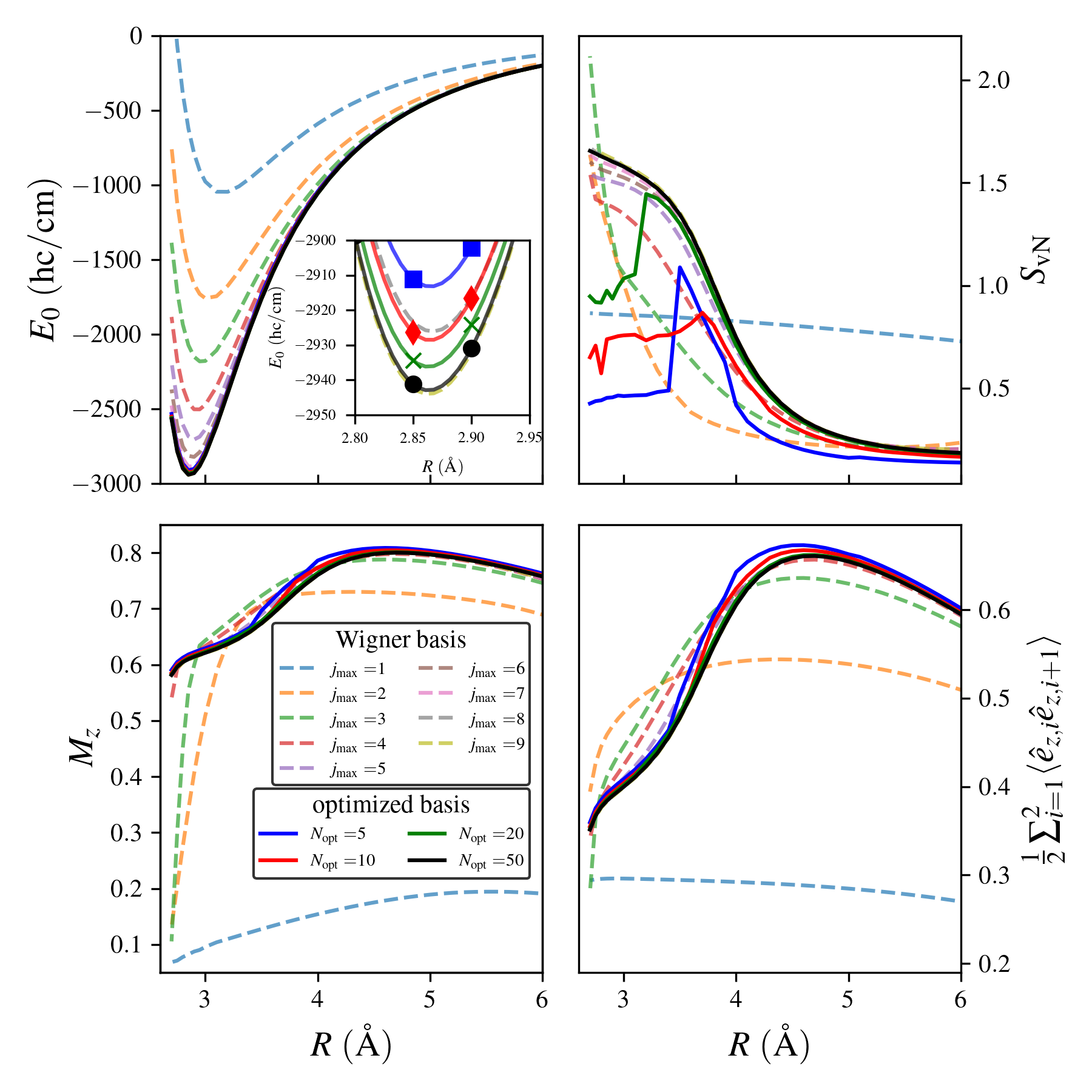}
\caption{Energy (upper left), von-Neumann entanglement entropy (upper right), polarization (lower left), and correlation function (lower right) calculated for a chain with $N=3$. Both the Wigner basis and the localized basis are used where the latter uses $j_{\mathrm{max}}=9$.}
\label{fig:R_scan}
\end{figure}

\begin{figure}
\centering
\includegraphics[width=\columnwidth]{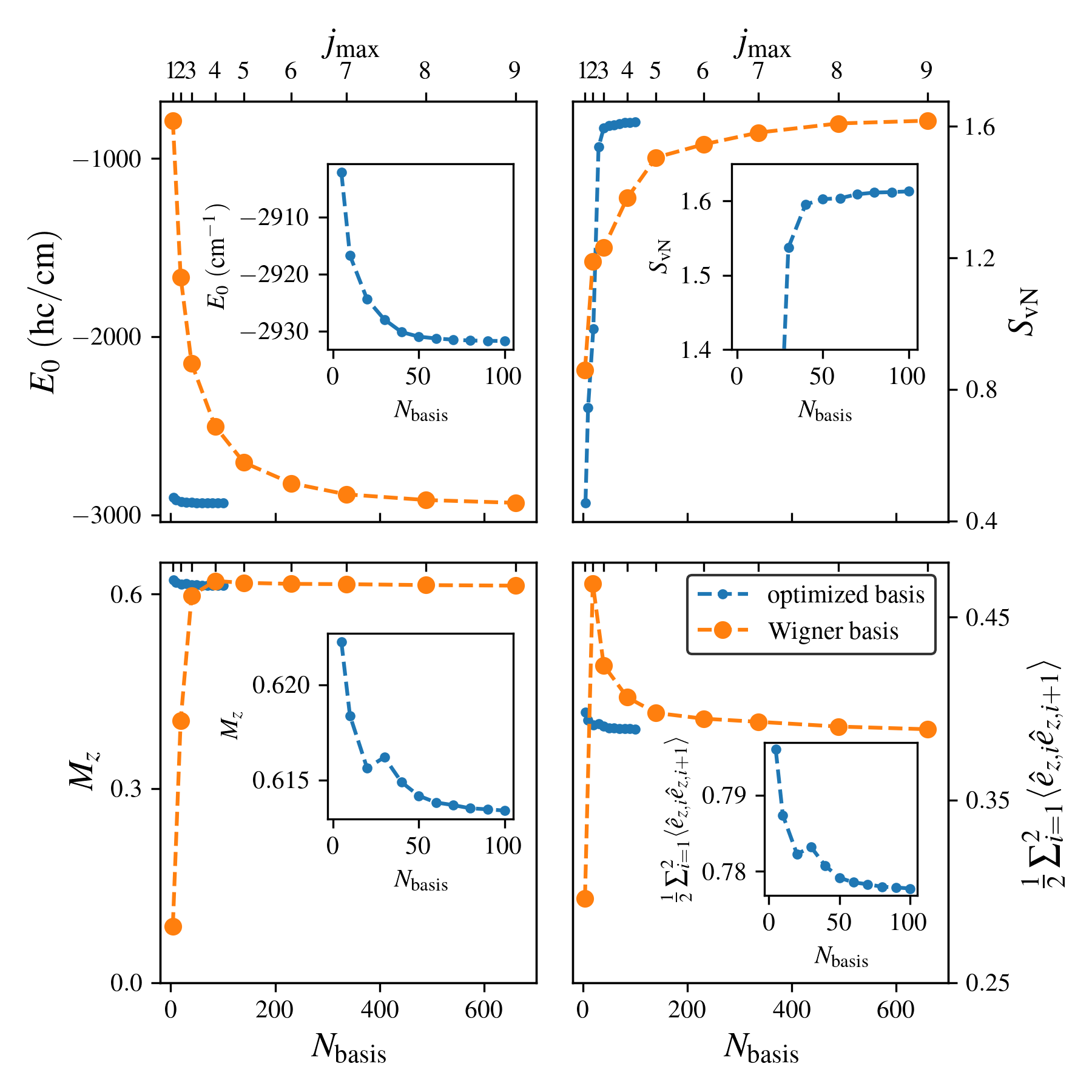}
\caption{Energy, von-Neumann entanglement entropy, polarization and correlation function calculated for a chain with $N=3$ at $R=\SI{2.9}{\angstrom}$. All properties are calculated with the Wigner basis and the localized basis where the latter uses $j_{\mathrm{max}}=9$. The system size is $N=3$.
}
\label{fig:R29}
\end{figure}

We begin our discussion with the water trimer because on the one hand, a large Wigner basis can be used to obtain a converged ground state since the system is small. 
On the other hand, the trimer is large enough to exhibit all the characteristics of a chain. Indeed, the system contains edge molecules with only one neighbor along with an inner water molecule with two nearest neighbors, and therefore provides a representative benchmark. We start by calculating several properties for various $R$ values with the optimized basis and compare with the corresponding values obtained in the Wigner basis. The results are presented in Fig.~\ref{fig:R_scan}.
For the interaction energy, the results shows that a small number of optimized functions is sufficient to converge the ground state. With only five optimized functions, it is possible to obtain almost \SI{99}{\percent} of the energy and with 50 optimized functions the energy is basically converged as shown in the  inset of Fig.~\ref{fig:R_scan}. This behavior holds for all $R$ values, from regions with purely dipolar interactions as well as for regions where hydrogen bonding occurs. Comparing the optimized basis to the Wigner basis reveals a convergence behavior of the former that is an order of magnitude faster than that of the latter basis set. 
As already mentioned, the energy is almost converged with five optimized functions. This number is comparable to the Wigner basis with $j_{\mathrm{max}}=1$ (dashed blue curve in Fig.~\ref{fig:R_scan}). However, this basis yields much higher energies that are far from being converged. In order to obtain a global accuracy comparable to the five optimized functions almost 500 hundred Wigner basis functions ($j_{\mathrm{max}}=8$) have to be used as seen in the inset of Fig.~\ref{fig:R_scan}. The supremacy of the optimized basis is most pronounced in the hydrogen bonding region where the water molecules are strongly localized. The convergence behavior for a fixed $R$ in that region is depicted in Fig.~\ref{fig:R29} and illustrates the much faster convergence behavior of the optimized basis compared to the Wigner one. 

Apart from the energy, Fig.~\ref{fig:R_scan} shows that other properties like the axial polarization $M_{z}$ and the nearest-neighbor correlation can also be converged with a relatively small number of optimized functions. Again, with only five optimized functions the respective curves already possess all the characteristic features (maxima, inflection points) whereas employing the Wigner basis with $j_{\mathrm{max}}=1$ yields curves that are completely off. 
In particular, the hydrogen bonding region at around \SI{3}{\angstrom} (see also the insets in Fig.~\ref{fig:R29}) and the purely dipolar region when $R<$ \SI{5}{\angstrom} are almost converged with five optimized basis functions. 
Between those distance values, at least 20 optimized functions are necessary to get a comparable accuracy. 
From previous studies, it is known that this region marks the transition between purely dipolar interactions and hydrogen bonding.\cite{sahoo2021path,serwatka2022ground} This region is characterized by the appearance of hydrogen bonds while there is still some rotational freedom of the individual molecules. 
A larger number of basis functions is therefore necessary to satisfy the dipolar and hydrogen bond character of the interaction.

The energy, polarization and the correlation function are properties that are obtained from integrals of the wave functions over the configurational space. 
To asses the quality of the optimized basis, we also present in Fig.~\ref{fig:R_scan}  the von-Neumann entanglement entropy because according to Eq.~\eqref{eq:SvN}, this property is directly derived from the wave function and should be even more sensitive to the quality of the computed state. 
Indeed, the entanglement entropy shows a slower convergence behavior than other properties.
This can be observed for distances above \SI{4}{\angstrom}. 
For $R<\SI{4}{\angstrom}$, small basis sets even lead to a wrong qualitative behavior as depicted in Fig.~\ref{fig:R_scan}. 
This slower convergence behavior of $S_{\mathrm{vN}}$ compared to the other three observables at $R=\SI{2.9}{\angstrom}$ is also depicted in Fig.~\ref{fig:R29}. The wrong qualitative behavior of $S_{\mathrm{vN}}$ for small basis sets is algebraic in nature since a certain minimal bond dimension is necessary to obtain a specific numerical value. But the maximal bond dimension a given MPS can reach is constrained by the local basis set size. 
If this size is too small to allow the necessary bond dimension, the correct entanglement entropy cannot be computed. This is an algebraic constraint and independent of the specific basis set. 
But if the size is large enough, the correct entanglement entropy can be obtained. 
Even though the convergence of $S_{\mathrm{vN}}$ is slower compared to that of the properties obtained from integral expectation values, $S_{\mathrm{vN}}$ can be also converged with fifty optimized basis functions.
This is a considerably smaller basis than the Wigner one. Thus, for different properties, the optimized basis shows a high ability to capture the correct physics and at the same time, provides a very efficient representation.

\begin{figure}
\centering
\includegraphics[width=\columnwidth]{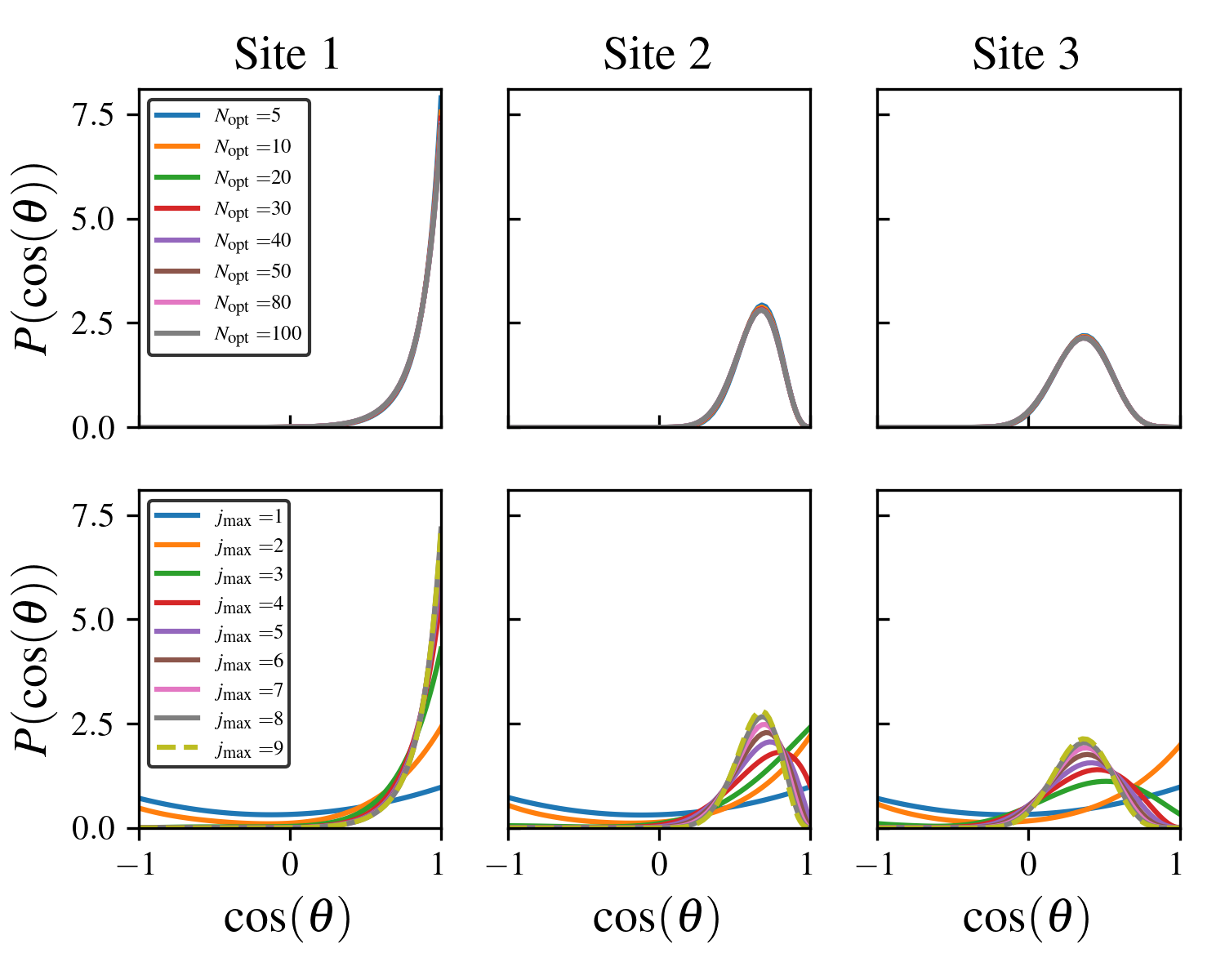}
\caption{Distributions along $\cos (\theta)$ of the different sites calculated in the localized basis (upper panel) and in the Wigner basis (lower panel). The distributions are calculated at $R=\SI{2.9}{\angstrom}$.}
\label{fig:Ptheta}
\end{figure}

\begin{figure}
\centering
\includegraphics[width=\columnwidth]{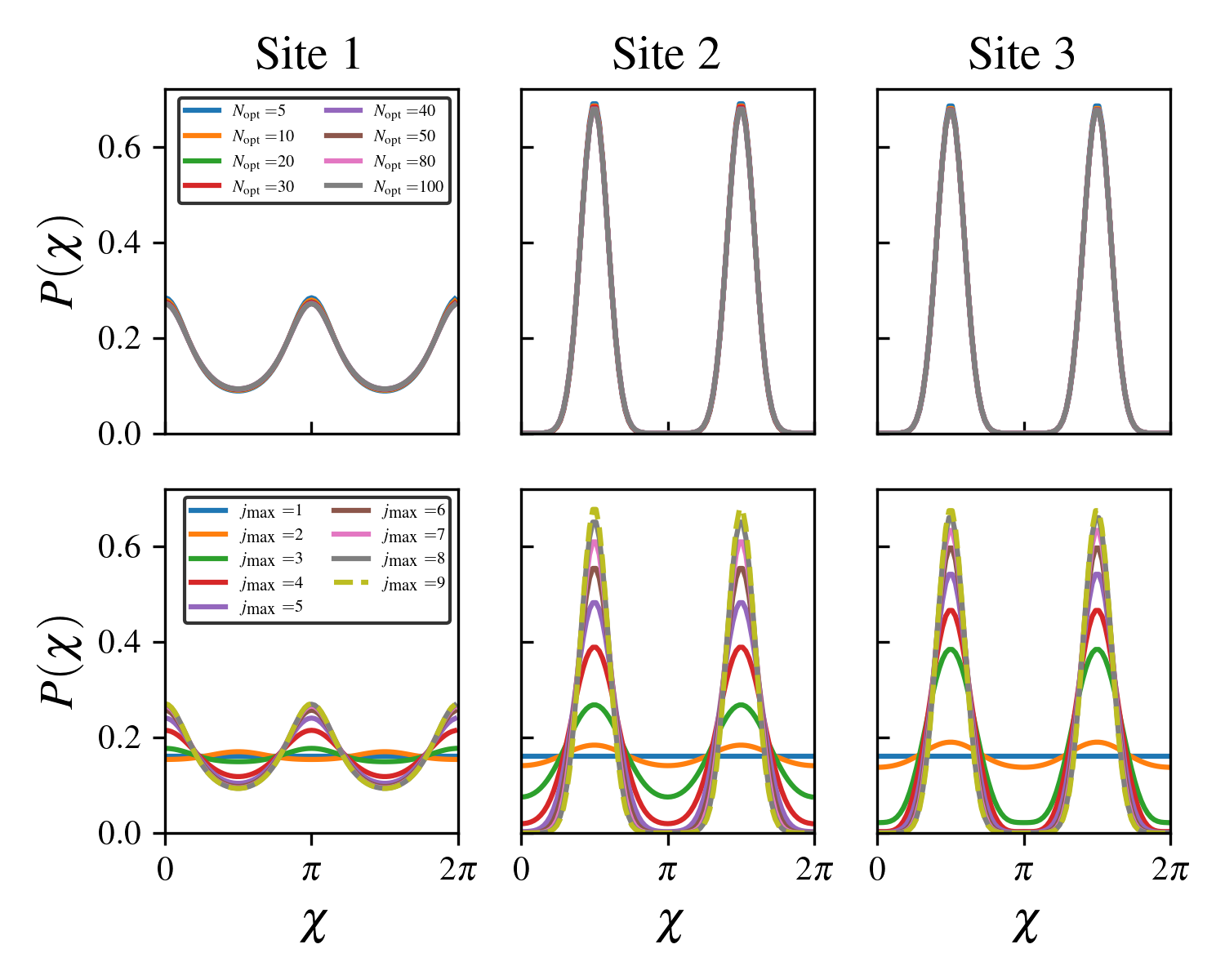}
\caption{Distributions along $\chi$ of the different sites calculated in the localized basis (upper panel) and in the Wigner basis (lower panel). The distributions are calculated at $R=\SI{2.9}{\angstrom}$.}
\label{fig:Pchi}
\end{figure}

\begin{figure}
\centering
\includegraphics[width=\columnwidth]{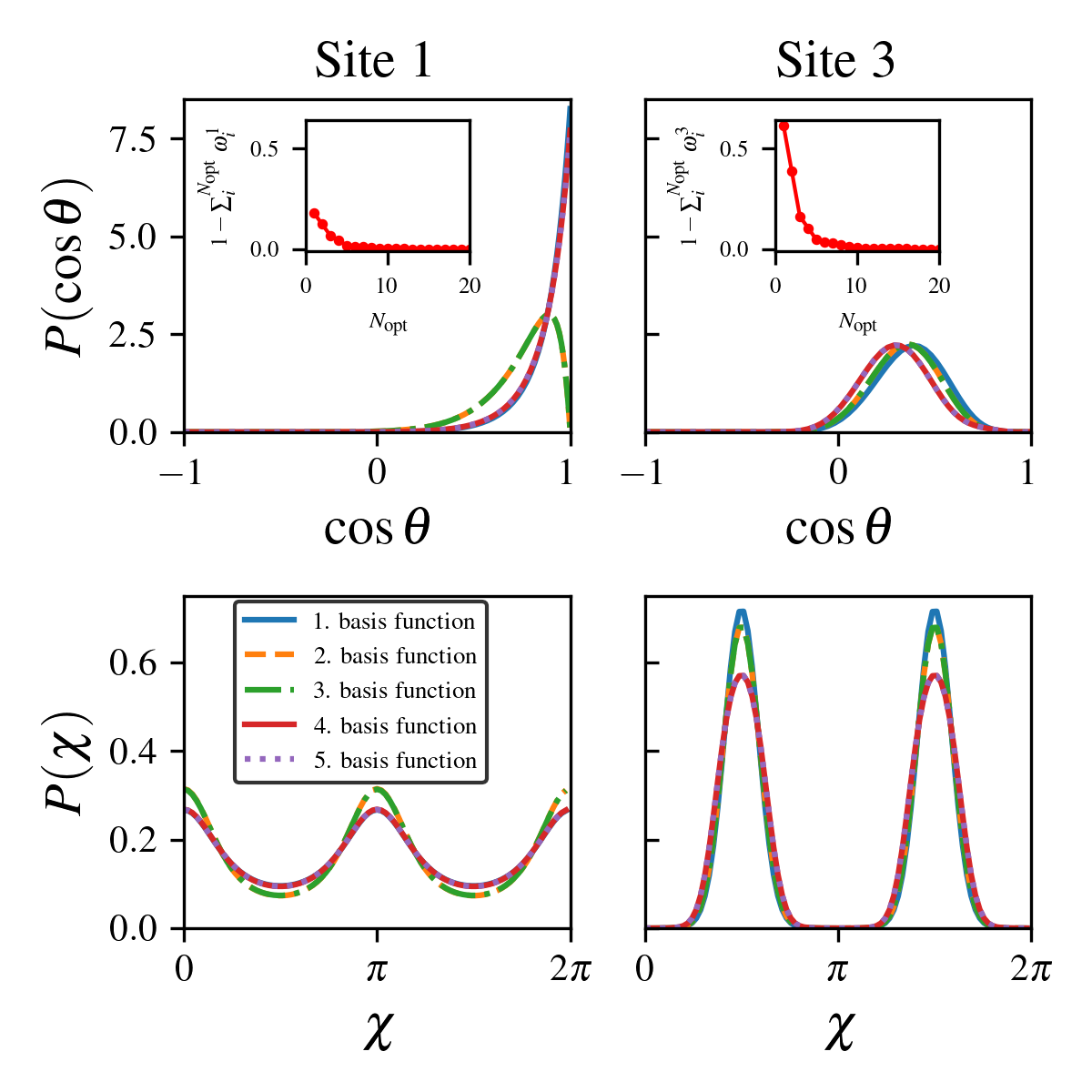}
\caption{Upper panels: angular distributions along $\cos (\theta)$ of the  first five localized basis functions at site 1 and 3. Insets: cumulative eigenvalues of the reduced density operators. Panel 2-4: angular distributions of the  first five localized basis functions at each site. Lower panels: angular distributions along $\chi$ of the  first five localized basis functions at site 1 and 3.}
\label{fig:Basis_ditribution}
\end{figure}

In particular, strongly localized regions are well described by the optimized basis set as revealed by the angular distributions for the three water molecules depicted in Figs.~\ref{fig:Ptheta} and \ref{fig:Pchi}. 
These distributions, at a water-water distance of \SI{2.9}{\angstrom}, are calculated for every site while the remaining sites are integrated out. 
In Fig.~\ref{fig:Ptheta}, the distribution along $\cos (\theta)$ shows the occurrence of maxima around 0.7 for the last two water molecules. 
These maxima signal the tilting of the molecule's dipole axis due to the formation of hydrogen bonds. 
In the optimized basis, only five functions are sufficient to describe the $P(\cos(\theta))$ distributions almost perfectly. 
These functions contain all the components of the primitive $j_{\mathrm{max}}=9$ basis that are necessary to capture this high angular localization. 
In contrast, the Wigner basis requires many more functions. 
A Wigner basis with $j_{\mathrm{max}}=4$ (85 functions) is the minimum required to obtain the shifted maxima in the distributions at site 2 and 3 but these are still a bit off the true maxima. 
Converged maxima at site 2 and 3 require at least $j_{\mathrm{max}}=8$ which corresponds to almost a hundred times more functions than the number required for the optimized basis. 
The distribution at the first site converges faster because it is centered at around 1, a natural maximum of many Wigner functions. 
As similar convergence behavior as for $P(\cos(\theta))$ can be observed for the distribution along the azimuthal angle $\chi$ (see Fig.~\ref{fig:Pchi}).
Due to the periodicity of the distributions, the position of the maxima can be captured with a smaller Wigner basis for $P(\chi)$ when compared to the $P(\cos(\theta))$ case.
However, the number of basis function to converge the distributions is still at least one order of magnitude larger compared to the optimized basis.
The high ability of the optimized basis to match the angular distributions using a few basis functions can be further understood by looking at the angular distributions of the first five basis functions at site 1 and 3 as shown in Fig.~\ref{fig:Basis_ditribution}.
Along $\cos(\theta)$ and $\chi$, the optimized basis functions already lead to the correct shape of the ground state distribution. This is not the case for the Wigner basis. 
As explained in Sec.~\ref{sec:Obasis}, the great advantage of a density-based basis is the fact that the eigenvalue associated with each basis function can be used as a criterion to judge the importance of the respective function to the total wave function. 
Hence, the sum of those eigenvalues can provide a measure of the quality of the basis set. 
The errors of these sums, represented as $1-\sum_{k}\omega^{i}_{k}$, are plotted in the inset of the top panels of Fig.~\ref{fig:Basis_ditribution}. 
They exhibit a very fast decay such that only the first few functions contribute significantly.
This corresponds to  the rapid convergence behavior observed of various properties.

\begin{figure}
\centering
\includegraphics[width=\columnwidth]{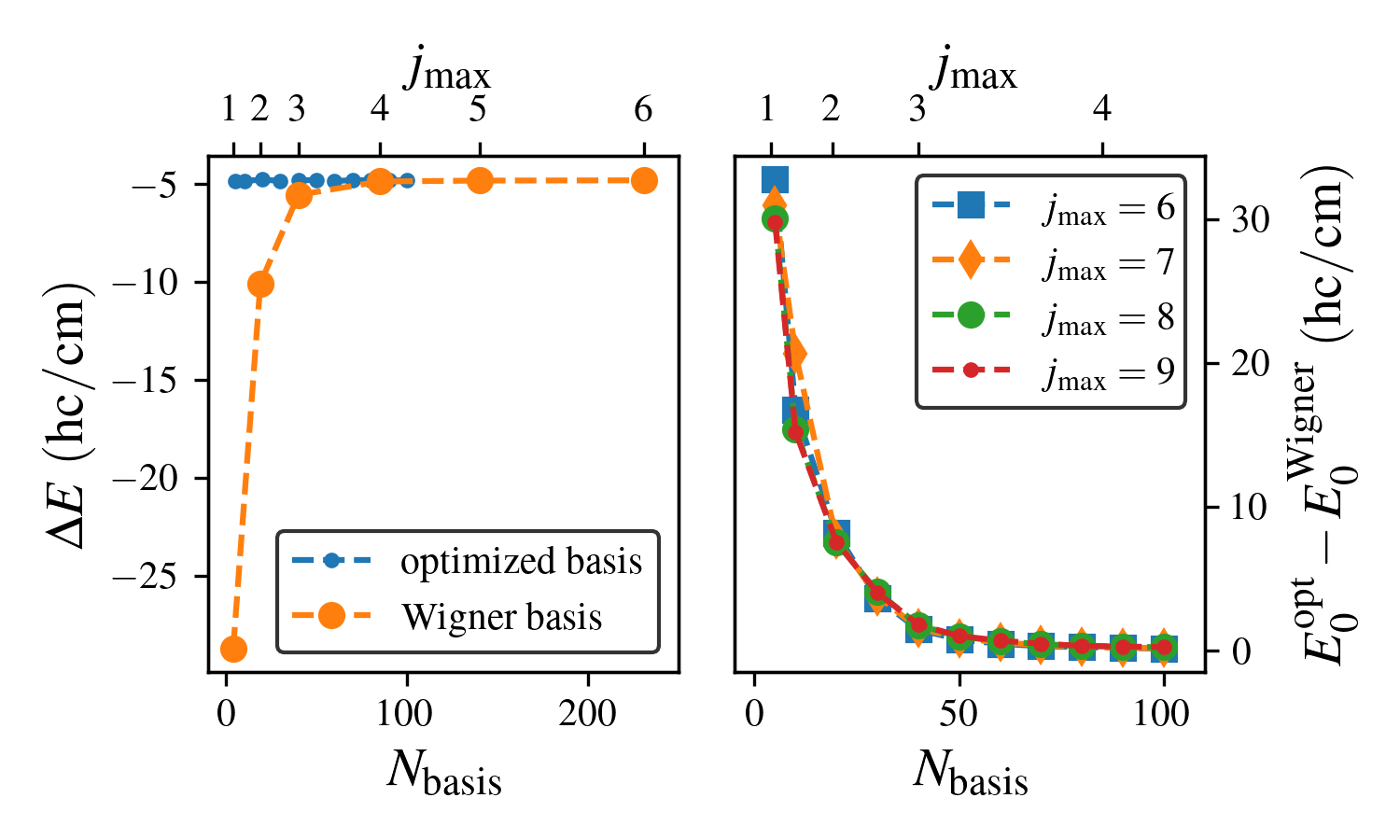}
\caption{Energy difference between the first two eigenstates calculated with the Wigner basis and the optimized basis (left panel). The latter uses $j_{\mathrm{max}}=9$. Energy convergence with respect to the number of basis functions for the optimized basis with different $j_{\mathrm{max}}$ values in the primitive basis (right panel). The system size is $N=3$.}
\label{fig:Egap_convergence}
\end{figure}

The quality of the optimized basis set, i.e. its ability to capture the relevant physics, depends on the underlying primitive basis. 
The primitive basis provides the pool from which the optimized basis can extract the components necessary to best describe the system. 
However, the larger the primitive basis, the higher the numerical effort required to compute the potential matrix in that primitive basis and the more DMRG steps have to be performed in one optimization iteration. 
In some cases it may be convenient to reduce the computational effort by using a smaller primitive basis. In Fig.~\ref{fig:Egap_convergence} we show the convergence behavior for the energy of a trimer at $R=\SI{2.9}{\angstrom}$ calculated by using different primitive basis set sizes ranging from $j_{\mathrm{max}}=6$ to $j_{\mathrm{max}}=9$.
The curves show the energy difference between the optimized basis and the Wigner basis for the respective $j_{\mathrm{max}}$. All curves obey a very similar convergence behavior. Independent of the primitive basis, $\sim\SI{99}{\percent}$ of the total energy can be obtained by only five optimized basis functions which corresponds to a total energy difference of about \SI{30}{\per\centi\metre}. For fifty functions and more, this error falls below \SI{1}{\per\centi\metre} and for one hundred functions, the energy is essentially converged. 
Hence, even if smaller primitive basis sets are employed, a few tens of optimized basis functions are sufficient to obtain converged results. The overall quality of the final results depends of course on the chosen primitive basis.

One issue that emerges from the use of a density matrix based optimal basis is a bias towards the ground state wave function. This makes such a basis set less suitable for excited states. 
However, for the system at hand, the strong bias of the localized basis can actually be exploited to calculate the first excited state. 
As we showed in previous work, for small symmetry breaking fields, the first excited state in water chains resembles the ground state but with all molecular dipoles inverted.\cite{serwatka2022ferroelectric,serwatka2023quantum} In order to capture this state in a DMRG calculation, we inverted the direction of the axial electric field in the initialization step. This yields an initial basis that points in a direction opposite to that of the previous set. 
Afterwards, for the following optimization steps, the field direction is switched back. 
But now, in the initial basis, the true ground state is shifted to much higher energies since we employ a truncated basis which is highly biased towards a state of opposite polarization. Basically, the opposite polarization of the initial basis penalizes the ground state and turns it into an excited state due to the finite size of the basis. In the subsequent optimization steps, this penalty increases and the DMRG calculation converges towards the first excited state of the true Hamiltonian. 
The fundamental energy gap of $\sim\SI{5}{\per\centi\metre}$ calculated by this procedure is shown in Fig.~\ref{fig:Egap_convergence}. 
As in the case of other properties, the gap converges very quickly. 
In comparison, the Wigner basis requires many more  functions. Interestingly, in the Wigner basis, the energy gap converges much faster than the total energy. That is probably due to the inverted but otherwise similar structure of the ground and first excited state which leads to a similar convergence behavior in both states and thus an error cancellation. Higher-lying states will have a more mixed character and are thus not suitable for the procedure just described. 
However, the combination of a left and a right-polarized basis functions to a new basis set might be a suitable way to describe more excited states.      

\subsection{Longer water chains}
\begin{figure}
\centering
\includegraphics[width=\columnwidth]{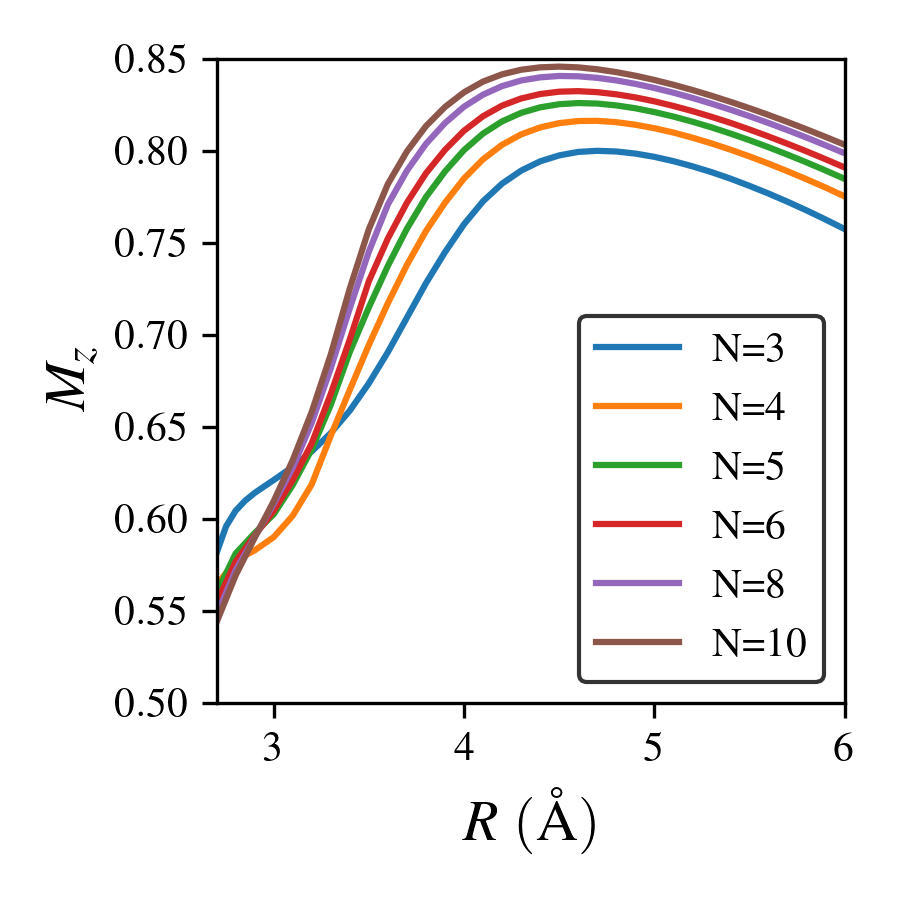}
\caption{Axial polarization for chains containing different numbers of water molecules.}
\label{fig:muz_nsites}
\end{figure}

\begin{figure}
\centering
\includegraphics[width=\columnwidth]{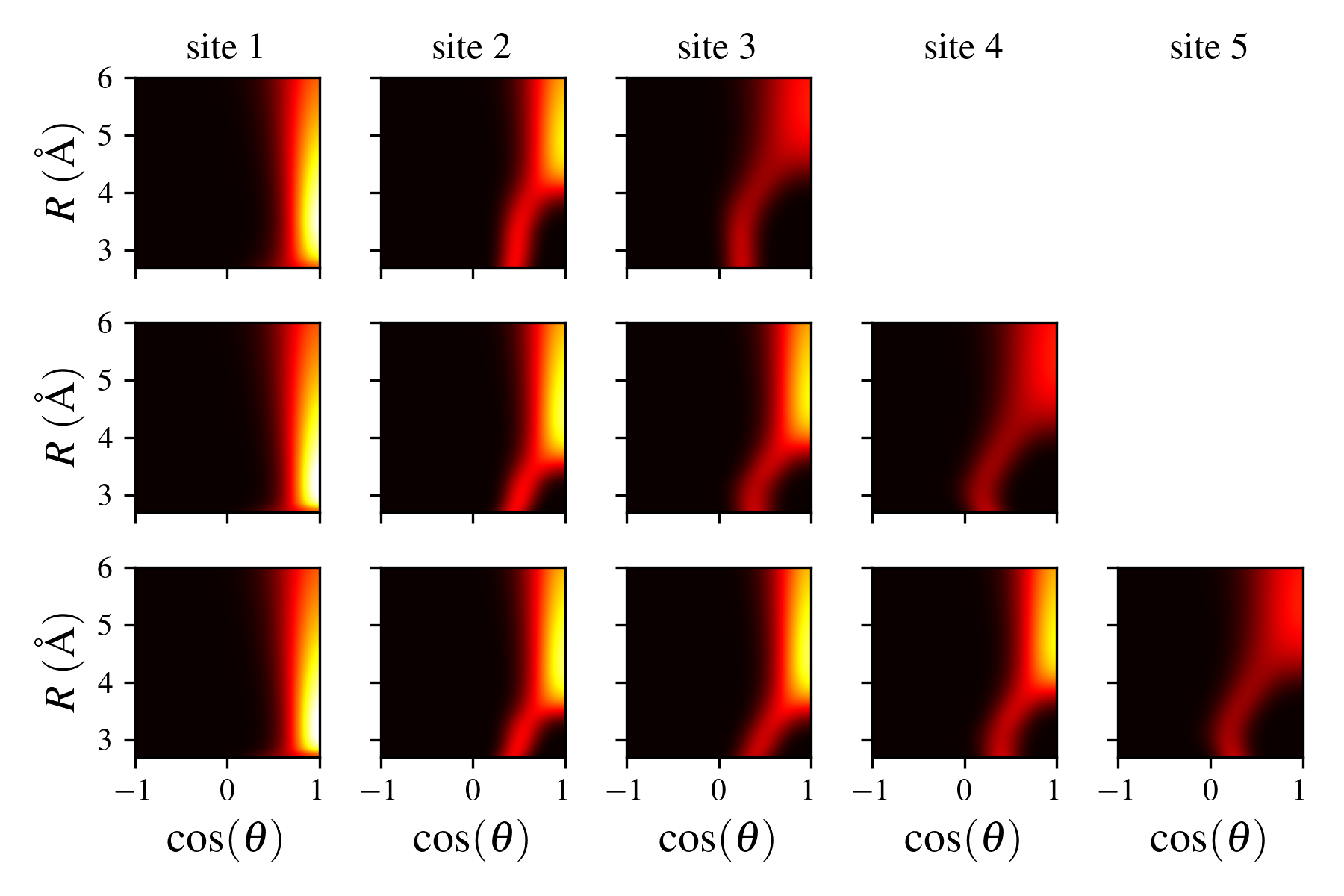}
\caption{Angular distribution $P(\cos(\theta);R)$ as a function of the distance parameter $R$ for chains with $N=3,4$ and $5$.}
\label{fig:pt_nsites}
\end{figure}

\begin{figure}
\centering
\includegraphics[width=\columnwidth]{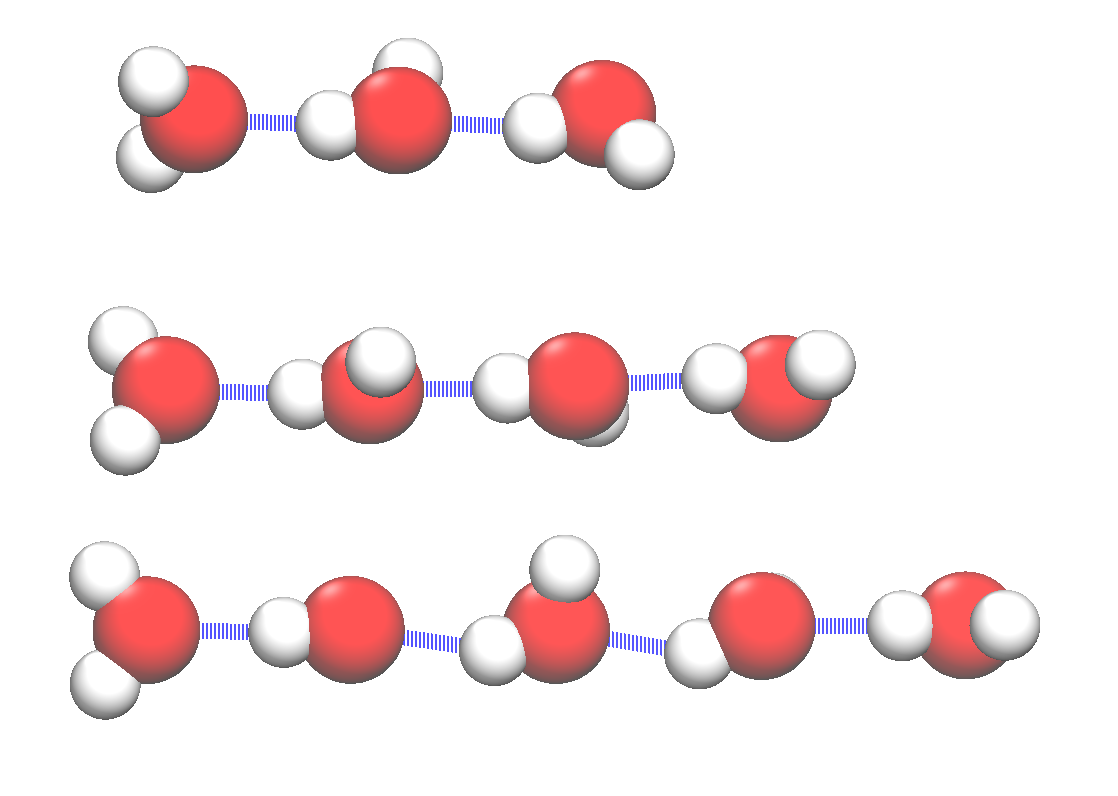}
\caption{Typical configurations for water chains with $N=3,4$ and $5$.}
\label{fig:chains}
\end{figure}

We finally present an application of the optimized basis to longer chains of up to ten water molecules. 
For these chains, the axial polarization $M_z$, is calculated and presented in Fig.~\ref{fig:muz_nsites}. 
Irrespective of the number of water molecules, all curves in Fig.~\ref{fig:muz_nsites} exhibit the same qualitative behavior. They show a growing polarization for decreasing $R$ up to a maximum value between 4.0-\SI{4.5}{\angstrom}. For smaller $R$ values, the polarization decreases but there is still a considerable net polarization of more than one half of the maximal (classical) possible value. Hence, even in the equilibrium region at around $R=\SI{2.9}{\angstrom}$, the linear water chain exhibits ferroelectric behavior regardless of the number of water molecules. 
This verifies a recent study in which a reduced computational setup was employed to show the non-vanishing polarization of a water chain embedded in a (6,5)-carbon nanotube at the equilibrium distance.\cite{serwatka2022ferroelectric} 
The existence of a maximum polarization can be explained by a change in the bonding arrangement from purely dipolar interactions to hydrogen bonding at shorter distances. 
To explain the quantitative differences between the curves in Fig.~\ref{fig:muz_nsites}, and in order to look at the hydrogen bond formation in more details, it is instructive to inspect the one-site polar densities for varying $R$ values as depicted in Fig.~\ref{fig:pt_nsites}.
Based on their behavior at small $R$, one can identify three types of water molecules. 
The first type is a pure hydrogen acceptor. For the chosen polarization of the external field, the left edge water molecule (site 1) belongs to this type, as can be seen in a classical representation of the chains depicted in Fig.~\ref{fig:chains}.
This acceptor molecule is localized at around \SI{0}{\degree} in the whole range from 2.7-\SI{6.0}{\angstrom}. Figure~\ref{fig:pt_nsites} shows that the localization gets stronger with decreasing $R$. 
This is caused by the increasing interactions that restrict the molecule's freedom to tilt away from the chain axis. 
At the right edge of the water chains, a second type can be observed, a pure hydrogen donor. 
In the purely dipolar region, ($R>\SI{4.5}{\angstrom}$) this configuration is localized at around \SI{0}{\degree} as all water molecules are.\cite{serwatka2022ferroelectric} 
But for shorter distances, that water molecule tilts to form a hydrogen bond with its neighbor. 
This tilting can be seen as a shifted distribution centered at around $\sim$\SI{70}{\degree} in Fig.~\ref{fig:pt_nsites}. 
The third configuration type includes all inner water molecules. 
All these molecules have two nearest neighbors and therefore act as hydrogen acceptors as well as donors. 
Like the right edge water, these molecules have to tilt in order to form hydrogen bonds as seen from the shift in the distributions in Fig.~\ref{fig:pt_nsites}. 
However, the two neighbors per water restrict its motion more strongly. 
Therefore, compared to the right edge water, smaller distances with stronger interactions are necessary to initiate the formation of hydrogen bonds which can be seen from the fact that the shift of the distributions occurs at shorter distance. Within the group of inner molecules, the shifts starts at slightly larger $R$ for the outer water molecules which experience less neighbors in total (compare site 2 and 3 for $N=4$ or site 3 and 4 for $N=5$). 
The larger rotational restriction of the inner molecules leads to smaller tilt angles which lie between \SI{45}{\degree} and \SI{54}{\degree} for the chains considered here. 
The distributions shown in Fig.~\ref{fig:pt_nsites} can be connected to the curves in Fig.~\ref{fig:muz_nsites}. The occurrence of the maximum can be easily connected to the start of the shift of the polar distributions caused by the formation of hydrogen bonds. 
The quantitative differences between the curves in Fig.~\ref{fig:muz_nsites} can be explained by the relative contributions of the three water types. With increasing $N$, the relative impact of the edge molecules decreases. 
Since the pure hydrogen donor starts to form hydrogen bonds earlier than the inner molecules, its declining importance leads to a polarization maximum that shifts to smaller $R$ for larger $N$. The rising impact of the inner water molecules with their smaller tilt angle also leads to a larger polarization with increasing $N$. 
It is also interesting to note, that the water molecules in the calculated chains prefer the formation of single hydrogen bonds instead of bifurcated bonds as can be seen in Fig.~\ref{fig:chains}. 
This is in agreement with a previous path integral ground state study where it was shown that the single hydrogen bond is energetically more stable than the bifurcated arrangement.\cite{sahoo2021path}

\section{Concluding Remarks}\label{sec:conclusio}
We employed an iterative approach by Zhang, Jeckelmann and White to construct optimized basis functions based on the reduced density matrix to calculate ground states of chains of rotating water molecules.\cite{zhang1998density} 
In this approach, the population eigenvalues of the one-body basis functions provide a clear truncation criterion. 
This allows one to assess the importance of the respective basis functions for construction of the the total state. 
A detailed convergence study of the water trimer showed that for most properties such as the energy or the polarization, only five basis functions per site are already sufficient to obtain nearly converged results. 
Only the entanglement entropy, a property that is directly derived from the wave function itself, requires a larger basis. But with a slightly larger basis set of $\sim 50$ functions per site, all properties are essentially converged. Compared to the primitive Wigner basis, this reveals a reduction of the basis set by at least a factor of ten. 
We showed that by employing a biased initialization step in the optimization procedure, it is even possible to calculate the first excited state in the ordered phase that has a polarization opposite to that of the ground state. 
A combination of the basis sets of left- and right-polarized states might be a good basis to obtain further excited states. Finally, we used the optimization procedure to calculate the ground states of longer water chains. The axial polarization of these chains exhibits a maximum in the region of $R=4.0-\SI{4.5}{\angstrom}$ which is due to the change of the interactions from purely dipolar to hydrogen bonds. The formation of the latter was studied by means of angular distributions revealing the occurrence of different types of hydrogen donors and acceptors.

In future work the proposed optimization procedure can be employed to further study the equilibrium region of water chains. This would allow one to obtain the zero Kelvin equation of state of water in one dimension, a currently unsolved problem. Furthermore, that procedure can be extended to additional degrees of freedom and  may provide a way to incorporate translations and vibrations into DMRG calculations of molecular chains.     

\section*{Acknowledgements}
The authors acknowledge the Natural Sciences and Engineering Research Council (NSERC) of Canada , the Ontario Ministry of Research and Innovation (MRI), the Canada Research Chair program (950-231024), and the Canada Foundation for Innovation (CFI). T. S. acknowledges a Walter-Benjamin funding of the Deutsche Forschungsgemeinschaft (Projektnummer 503971734). 

\section*{DATA AVAILABILITY}
The data that support the findings of this study are available from the corresponding author upon reasonable request.

%

\end{document}